# Investigating the Perceived Impact of Maternity on Software Engineering: a Women's Perspective

Larissa Rocha
*State Uni. of Feira de Santana*
Feira de Santana–Bahia, Brazil
larissa@ecomp.uefs.br

Edna Dias Canedo
*University of Brasilia (UnB)*
Brasília–Distrito Federal, Brazil
ednacanedo@unb.br

Claudia Pinto Pereira
*State Uni. of Feira de Santana*
Feira de Santana–Bahia, Brazil
claudiap@uefs.br

Carla Bezerra
*Federal University of Ceara (UFC)*
Quixada–Ceara, Brazil
carlailane@ufc.br

Fabiana Freitas Mendes
*University of Brasilia (UnB)*
Brasília–Distrito Federal, Brazil
fabianamendes@unb.br

*Abstract*—Background: Several researchers report the impact of gender on software development teams, especially in relation to women. In general, women are under-represented on these teams and face challenges and difficulties in their workplaces. When it comes to women who are mothers, these challenges can be amplified and directly impact these women's professional lives, both in industry and academia. However, little is known about women ICT practitioners' perceptions of the challenges of maternity in their professional careers. Objective: This paper investigates mothers' challenges and difficulties in global software development teams. Method: We conducted a survey with women in the ICT field who work in academia and global technology companies. We surveyed 141 mothers from different countries and employed mixed methods to analyze the data. Results: Our findings reveal that women face sociocultural challenges, including work-life balance issues, bad jokes, and moral harassment. The prejudices they suffer make them insecure and with low confidence in the work performed. Furthermore, they usually do not have a supporting network during and after maternity leave, which culminates in them feeling overloaded. The surveyed women suggested a set of actions to reduce the challenges they face in their workplaces, such as: creating a code of conduct for men and childcare within companies. Conclusion: Women face many challenges when they become mothers. Our findings explore these challenges and can help organizations in developing policies to minimize them. Also, it can help raise awareness of co-workers and bosses, toward a more friendly and inclusive workplace.

## I. INTRODUCTION

There is a worldwide effort to increase gender diversity in Science, Technology, Engineering, and Mathematics (STEM), most notably in Information and Communications Technology (ICT) [1]. However, the representation of women in these areas is still low [2], [3]. Canedo et al. [4], and Izquierdo et al. [5] found that women account for approximately 10% of software development teams and are primarily under-represented in leadership positions. On the other hand, National Center for Women & Information Technology (NCWIT) Scorecard shows trends in girls' and women's participation in computing in the United States (U.S.) over the time. The above mentioned report shows that in 2019 women accounted for approximately 20.9% of students completing bachelor's degrees in ICT in the U.S. against 16.7% in 2009 [6].

Numerous research studies have been conducted in the context of gender diversity in recent decades, which indicate that women in ICT community have faced a quarrel culture on behalf of men that manifests itself as under-representation, disfavor, and inequality [4], [7]–[9]. The sociocultural challenges women generally face, whether in academia or the software industry, may ultimately drive them to quit jobs, mainly when gender diversity is not a priority in their organizations [10]. Kuechler et al. [11] identified that women's distancing from their jobs is related to them not being aligned with their motivations or due to unpleasant and hostile social dynamics in their workplace.

Women are susceptible to getting pregnant. Thus, they can be single mothers, have very young children that require constant care, be the family's sole provider, or deal with numerous family problems. These scenarios were even more evident during the COVID-19 pandemic when women working in ICT across different roles had to deal with at least two types of problems, one due to the pandemic scenario and the other due to the continued lack of consideration from male co-workers [12].

During the COVID-19 pandemic, software development teams were impacted by the migration from face-to-face work to remote work [13]–[15]. In such a scenario, mothers who were part of software development teams reported that working remotely with children was highly challenging, often carrying out their activities outside working hours [12]–[14], [16]. Furthermore, it is known that historically the responsibility for household and maternity activities is attributed to women, causing them to have a double shift and suffer penalties in the labor market.

Trinkenreich et al. [17] identified possible actions that companies in the ICT sector can apply in parental and gender-inclusive policies, such as: sponsoring child care, providing adequate maternity leave beyond the relevant country's laws, and also providing more flexibility in work hours. The same authors also reported that some challenges for women in the software industry are that mothers receive [fewer] responsibilities because they have kids. Also, when returning from maternity leave, the company does not provide enough support, which usually leads the woman to ask to step down from the role.

Despite numerous studies investigating gender in the software industry, we have not identified studies that deeply investigate the challenges of motherhood. Thus, this paper aims to investigate the challenges and difficulties mothers face in global software development teams. To achieve this, we address the ephemeral and critical challenges experienced by women in ICT working in either public organizations (e.g., educational, development, and research activities) or private software development companies.

Hence, we performed a survey with 141 mothers from 17 different countries. The survey was composed of 55 questions, 47 closed-ended and 8 open-ended questions. Our findings reveal that mothers face many challenges in their workplace, such as moral harassment, lack of empathy/sorority, work flexibility, and a support network. As a consequence, they feel overloaded, lack confidence, and are physically and mentally stressed, especially when they are single mothers and providers. It is exhausting for mothers to ask for and explain absences due to maternity, which can be many and of various types (e.g., doctor's appointments, kids' illnesses, and school events). Companies lack empathy, both from co-workers and bosses.

In addition, the main suggestions to mitigate the challenges pointed out by the surveyed mothers were: changing culture in organizations, creating a code of conduct for men, creating childcare within companies, creating opportunities/programs for women in ICT, and switching to hybrid work when needed.

## II. BACKGROUND AND RELATED WORK

### A. Impact of gender on software development teams

Several studies in the literature have investigated the impact of gender on software development teams, especially the participation of women [18]–[20]. Understanding the causes of this under-representation helps us understand the reasons for gender imbalance, whether in academia or industry. In addition, it can help organizations devise strategies to attract and retain women in ICT.

Wolff et al. [21] surveyed 252 women, and they reported a lack of self-efficacy, which is a potential predecessor of imposter syndrome. The authors reported that women feel discriminated against regarding equal opportunities, and their negative experiences in their workplace environments may affect their feelings and attitudes toward their careers. Canedo et al. [4] have conducted semi-structured interviews with 17 Brazilian women, and they reported dealing with hostile sexism, including discrimination and prejudice, benevolent sexism (i.e., when women are not given the most complex tasks to perform), and glass ceiling, whereby few women hold leadership roles in their teams.

Women contributing to OSS projects also reported work-related challenges, including work-life balance issues and imposter syndrome [10], lack of parity with colleagues and sexism [7], and prove-it-again [22]. Some workers also identified that non-inclusive communication is faced by women contributing to OSS projects, particularly technical biases against women developers with lower code acceptance rates, as well as delayed feedback during code reviews and discussion lists [23]–[25]. Women also reported hostile sexism they faced during meetings with contributors.

### B. Maternity in software engineering

In the literature, we did not identify studies that directly investigated the impact of motherhood in the Software Engineering context. However, some studies report findings about challenges that mothers face in the software industry [4], [5], [7], [12], [17], [22]. During the transition to remote work due to the COVID-19 pandemic, Bezerra et al. [13] and Machado et al. [26] conducted a study with Brazilian women. They reported that women faced even more work-life balance problems during the pandemic, lacking support in performing household chores and taking responsibility for their children.

Unlike previous work, this paper explores the difficulties and challenges that mothers experience in software development teams and academic environments. In addition, we also aim to identify suggestions for dealing with these challenges.

## III. METHODOLOGY

This study aims to investigate the impacts and challenges of motherhood in Software Engineering, both in academia and industry. In order to reach this goal, we defined the following research questions (RQs):

RQ.1: **Do mothers remain employed after becoming pregnant? If they quit their jobs, what were the motivations?** This question aims to identify the main reasons why ICT women leave their jobs after becoming mothers.

RQ.2: **How was the maternity leave affected by the job activities?** This question investigates whether mothers had to work during the leave or if they had to anticipate returning to work.

RQ.3: **What were the strategies employed by mothers to overcome difficulties during the COVID-19 pandemic?** The COVID-19 pandemic may have impacted the balance between work and motherhood activities. This question seeks the set of strategies that mothers used to make work possible during the pandemic.

RQ.4: **What are the harassment and prejudice related to motherhood that women face while working in the ICT area?** ICT area is predominantly male context and women may suffer many prejudices related to maternity. This question details the type of harassment or prejudice that women suffer after becoming mothers.

RQ.5: **What are the perceived difficulties by mothers who occupy positions within the computing area?** This question aims to unveil the main difficulties mothers face in ICT jobs, both in industry and academia.

RQ.6: **What do mothers suggest to mitigate the difficulties faced at work related to motherhood?** This question aims to identify what actions mothers suggest to make the work environment more friendly.

The next sections detail the process of creating the survey, including the design and procedures to define the target audience, the pilot study, the survey invitation and distribution, and the strategies we employed to analyze the gathered data.

### A. Target Audience

We considered software practitioners women who are working in ICT and are mothers (pregnant, biological mother, stepmother, and adoptive mother). Hence, we included a control question at the beginning of the survey to filter the respondents and ensure we got responses only from our target audience.

### B. Survey Design

All the authors of paper were involved in the design and validation of the survey questions. Three authors described the survey questions and the others validated them. The survey consisted of 55 questions, 47 closed-ended and 8 open-ended questions grouped into 13 sections (S), as follows:

S1: Consent to participate in the research
S2: Control Question
S3: General Profile
S4: Children
S5: Professional Life
S6: Work in Industry
S7: Work in Academia
S8: Work in both Industry and Academia
S9: Organization
S10: Pregnancy
S11: Maternity Leave
S12: If not employed during maternity leave
S13: Difficulties, Challenges and Suggestions

At the beginning of the survey, we presented the statement of informed consent (S1), including the conditions and stipulations. We also presented the contact information. The survey was anonymous and respondents were not asked for any contact information, then, we had the control question (S2). The next section (S3) comprised the general profile with five questions; (S4) was related to children containing three questions; and (S5) asked about professional life through four questions.

From this point, we divided the survey into three sections, depending on the type of mother's job: software industry only (S6), academia only (S7), or both industry and academia (S8). Those sections had three, two, and five questions respectively.

Section (S9) contained three questions about the organization they worked for, common to the three jobs (industry, academia, or hybrid). The next four sections are related to maternity issues. (S10) consisted of nine questions about the pregnancy period; (S11) contained 14 questions about the maternity leave period; (S12) was aimed at those who were not employed when pregnant with four questions; and (S13) encompassed the last two questions regarding the maternity difficulties, challenges, and suggestions. The complete survey is available in our supplementary material[1].

### C. Pilot Study

We have conducted a pilot test round to evaluate the survey quality. We sent the questionnaire to three mothers who occupy positions within the computing area. Their feedback included suggestions regarding the wording of the questions and other modifications, such as including alternatives to closed questions and the researchers' contact information. We followed their advice and improved the questionnaire. Regarding the time to complete the survey, the pilot respondents, took less than 18 minutes on average. We reported that time to the respondents when the survey questionnaire was made public.

### D. Survey Invitation

We used the Google Forms platform[2] to create the survey questionnaire. Next, we made it available through cards and text on different social media platforms. We used two strategies: posts and direct messages. We posted on Twitter, LinkedIn, Facebook, and Instagram, and we sent direct messages to profiles on these platforms, in addition to WhatsApp and e-mails. The questionnaire was available from November 10th to December 14th, 2022 (34 days).

### E. Data Analysis

This research is essentially qualitative research using histograms and percentages to characterize the sample or rank the items most cited by the respondents. To answer the research questions, we employed elements of the Grounded Theory by performing open and axial coding [27]. Grounded Theory refers to a method of inductively generating theory from data. Studies often include unstructured text, for example, interview transcripts, field notes, and so on. However, they may also include structured text, diagrams, images, and even quantitative data [28].

In this study, the coding process was performed in three rounds. In the first one, two authors performed the open coding of all open questions. Thus, they split the data into discrete parts and labeled these parts to create codes. In the second round, the other two authors performed the axial coding. They read the discrete parts of the data and the assigned codes to identify connections among the codes and group them into categories. Finally, another author reviewed and refined the categories and codes in the third round. An example of the coding process is shown in Figure 1. The example shows respondent # R03's answer regarding the difficulties and challenges faced at work as a mother. The complete open and axial coding process is available in our supplementary material.

---
[1]Survey supplementary material available at https://zenodo.org/record/7548888
[2]https://www.google.com/forms

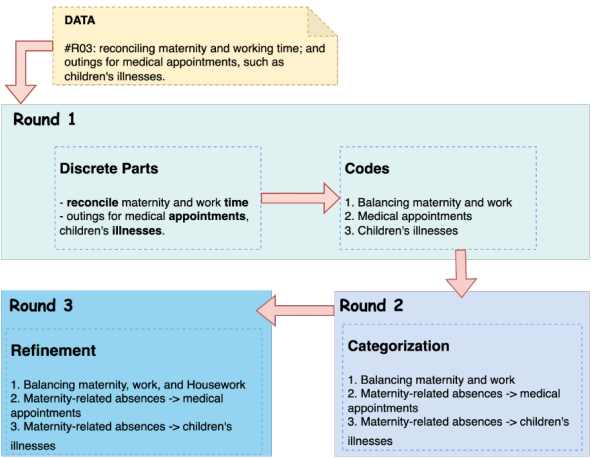

Fig. 1. Example of how the coding process was carried out.

## IV. THE MOTHERHOOD LANDSCAPE IN THE ICT AREA

The survey received 147 responses. After removing the responses from women who were not mothers (based on the control question), we obtained 141 valid responses.

### A. Respondents' profile

We received responses from 17 different countries. Overall, more than half of mothers work in Brazil (54.6%), but we also received answers from Portugal (6.4%), the USA (4.3%), Bolivia (4.3%), Belgium (3.5%), Canada (3.5%), Spain (3.5%), Argentina (2.8%), Sweden (2.8%), Australia (2.1%), Colombia (2.1%), Mexico (2.1%), Switzerland (2.1%), Austria (1.4%), Chile (1.4%), Denmark (1.4%), and United Kingdom (1.4%). Our research had the participation of mothers from different continents of the world, mitigating the cultural factor. Furthermore, mothers of different ages participated in the survey. In general, most respondents were between 31 and 54 years old (83.7%); 3.5% were 21 to 25 years old; 7.1%, 26 to 30; 2.8%, 55 to 60; and 2.8% were more than 61 years old.

Almost half of the respondents had 2 children (48.2%); 39% had only one child; 9.9%, had 3 children; 2.1%, 4 children; and 0.7% had more than 5 children. The survey questionnaire also asked about the age of the youngest child. For 63.8% of the mothers, their youngest child was up to 6 years old; for 12.8%, the youngest child was between 7 and 9; and for 23.4%, they were over 10 years old. Additionally, 97.2% of the respondents, live with their children. Most respondents have children aged up to 10 years. This factor helps the results of our research, as the age group needs more support and dedication from mothers.

Regarding education, only one respondent had not yet finished a bachelor's course. Of those who finished it, 36.2% pursued a Master's degree, 29.1% Ph.D., and 14.9% MBA. Of the total of respondents, 97.9% were employed when they answered the survey, and among them, 42.6% had a face-to-face job; 33.3% was working in a hybrid mode (remote and face-to-face); and 22% were working fully remote. For 42.6% of them, the family income was higher than 10 minimum wages. However, 11.3% received up to 3 minimum wages; 24.1% up to 5; 12.8% up to 7; and 9.2% up to 9 minimum wages. We also obtained in the survey representativeness of approximately 58% of the respondents working in a hybrid or remote way and 42% face-to-face. In addition, most respondents have a family income of more than 7 minimum wages.

More than half of the respondents were working at private software development companies (56%); while 32.7% of them were working at either Federal or State Public Administration; 14.2% at research/collaboration projects; 5.7% at State-owned companies (such as state banks); and 0.7% were working with open source software projects. We can see that most of the respondents are from the industry, but we also have a good representation of women who work in academia. The percentage is higher than 100% because 12 respondents marked more than one option. 53.9% of the women are married; 21.3% are single; 15.5% are in a long-term relationship; 7.1% are divorced; 1.4% are separated; and 0.7% are widowed. Thus, the general profile of the mothers who participate in the survey is Brazilian, with two children up to 6 years old, with a master's degree in computing, employed in a face-to-face or hybrid work regime, and receiving 10 minimum wages in a software development company.

### B. Organization

The respondents performed different roles in the organizations, such as Lecturer and Researcher, Software Engineer, Programmer/Developer, Project Manager, Requirements Analyst, Professor, Software Tester, Human-Computer Interaction specialist, Designer, and Researcher. Other roles, such as QA Tech Leader, Product Manager, Linguist, Infrastructure, Enterprise Architect, Engineering Manager, Director, Database Administrator, Data Modeling, Course Coordinator, Technology Coordinator, and Change Management Analyst, were mentioned only once by the participants. Figure 2 shows the main roles informed by them.

For those who work in the software industry, their team size is characterized as follows: 50% of respondents work in teams with more than 16 members; for 13%, the teams have from 11 to 15 members; for 23%, from 6 to 10; and for 14% less than 5 members. For respondents who work in academia, the numbers are much higher, since we considered the number of members in their departments. Thus, 20% of respondents work in departments with up to 20 members; 23%, from 21 to 49 members; 28%, from 50 to 99 members; and 29%, more than 100 members.

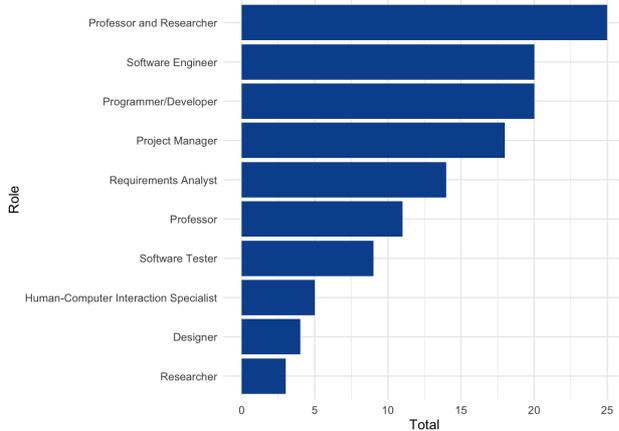

Fig. 2. Main roles performed in the organizations

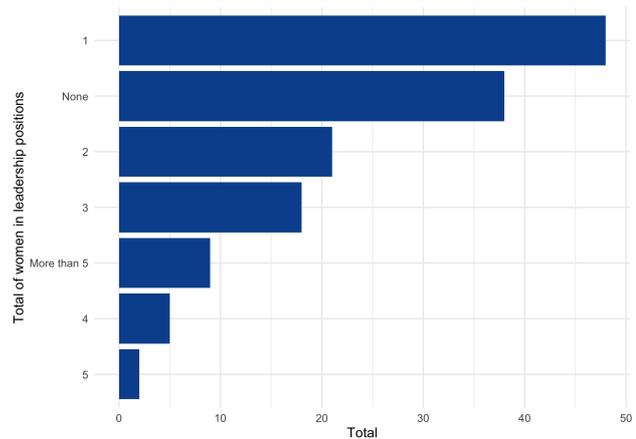

Fig. 4. Number of women in leadership positions in team or department

Although they occupy various roles in the software industry and academia and may work in huge teams or departments, 36% of respondents work in teams/departments with up to 6 women; 18% work with 7 to 15 women; 8% work with more than 16 women, and 57% of them affirmed that they work with less than 3 women, as shown in Figure 3. In addition, 61% of respondents stated that there is one or no women in leadership positions in their teams/departments. 27.7% stated that there are 2 or 3 women; 4.9% stated that there are 4 or 5; and 6.4% informed that there are more than 5 women in leadership positions in their teams or departments. This finding is similar to what was found by Canedo et al. [4] and Izquierdo et al. [5]. For 93.6% of respondents, work in teams with up to 5 women as leaders, as shown in Figure 4.

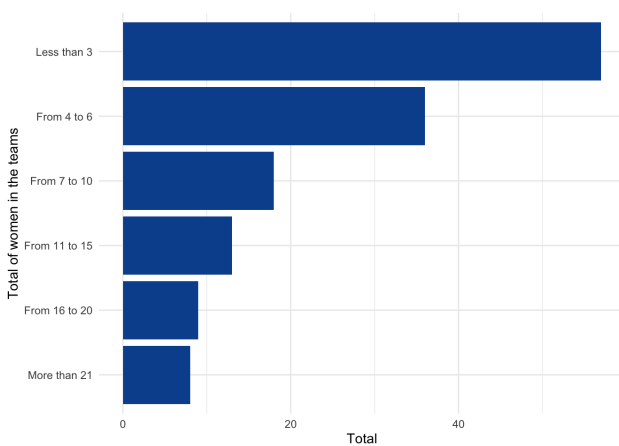

Fig. 3. Number of women in teams or departments

*1) Industry setting:* Of the total of mothers, 60.3% worked only in the software industry and 10.6% worked for Industry and Academia. For both groups, most respondents have worked in the industry for more than 15 years (34%), as Figure 5 shows. However, respondents were quite diverse, with some having little experience, such as less than 1 year (3%), to others between 4 and 9 years (23%), and even more than 10 years (65%).

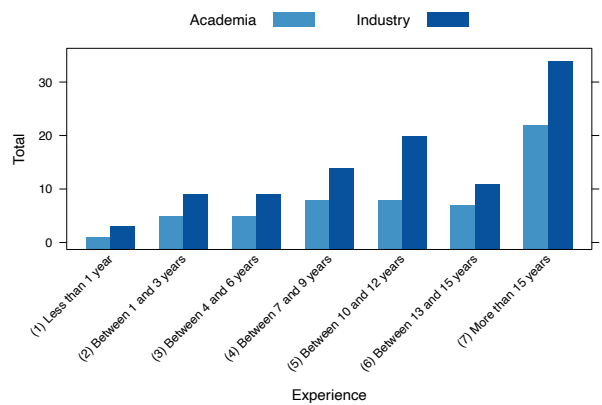

Fig. 5. Years of experience in the software industry and academia

Regarding the organization, 39% of the respondents work in large-sized companies with more than 400 employees, combining both groups, Industry and Industry & Academia, as Figure 6 shows. Yet, 33% work in medium-sized companies, with 199 employees; and 10% work in companies with up to 20 employees, considered as small or micro companies by the Organisation for Economic Co-operation and Development (OECD).[3]

Concerning the team size of women who work in Industry, 37% work in teams with more than 21 members; 40% work in teams with 6 to 20 members; and 8% work in teams with less than 5 people. Regarding the size of teams that women work on both in Industry and Academia, 4% work in teams with more than 21 members; 5%, from 6 to 15; and 6%, less than 5, as Figure 7 shows.

---
[3] classification available at: https://data.oecd.org

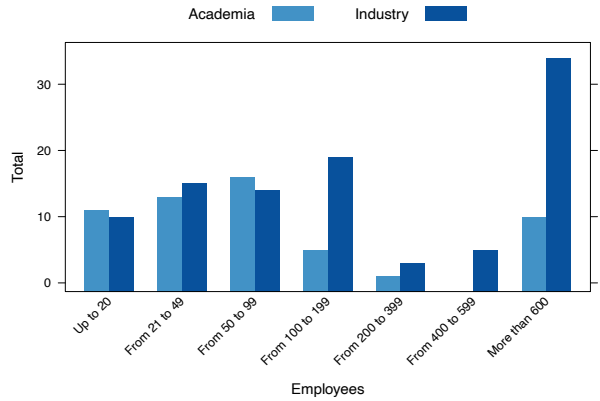

Fig. 6. Number of employees in the organizations

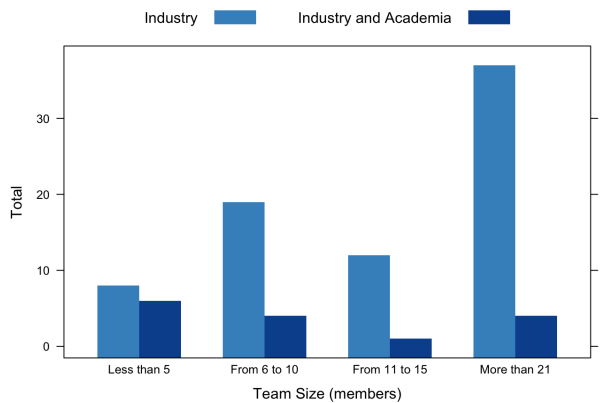

Fig. 7. Team size

*2) Academia setting:* Considering the total of respondents, 29.1% work exclusively in Academia, and 10.6% work for both Academia and Industry. For both groups, most respondents work in academia for more than 15 years (22%), as Figure 5 shows. However, respondents were quite diverse, with some having little experience, such as less than 1 year (1%), to others between 7 and 12 years (8%), and between 13 and 15 years (7%).

Regarding the respondents who work at the Academia, most of the respondents work in educational institutions with 50 to 99 employees (16%), 13% work in institutions from 21 to 49 employees, and 10% with more than 600 employees, as Figure 6.

## C. Reasons to quit or change jobs (RQ.1)

Considering the 141 respondents, 131 were employed when they last became pregnant, which is about 93% of them. From these 131 mothers, 14 (11%) left their jobs after they became pregnant. The answer of respondent #R62 is a warning about the problems suffered by women: *"In my first pregnancy [...] I was fired when I returned from maternity leave and it took me 2 months to find a new job."*

From the ones who indicated the reasons to quit or change their jobs, the main motivations were:

- Moral harassment (45.5%). They faced bad jokes, prejudice, and discrimination. For example, #R110 commented ***"I couldn't stand my co-workers, there were a lot of bad jokes and it lowered my self-esteem"***. Also, *"I looked for a job where people believed in me more"* (#R119); and *"I changed job because my boss didn't think I was capable of working and taking care of the kids"* (#R105);
- Health-related issues (18.2%). They had to deal with the health problems of both mother and baby. For example, respondent #R32 said *"my child was born with Down syndrome and this fact prevented me from working for two years and after this period, still struggling, I used to do my job during the evenings"*;
- Feeling guilty (9%). Reconciling work with motherhood is not an easy task. For example, respondent #R47 said *"we always feel guilty for going out and leaving the children at home with the nannies"*.

Although less cited, they also reported that changed their jobs because of logistics around child rearing and changed to a position that requires fewer hours and stress. Additionally, 27.3% of mothers reported that they left their jobs because they found better opportunities, for example, a better position or salary, some also reported that they switched to companies that allow remote work. For instance, respondent #R02, stated *"I had the desire to breastfeed my baby and the company did not allow remote work"*.

> **RQ.1 Summary**: 11% of the woman quit their jobs after becoming mothers. The main reasons for quitting work were: moral harassment suffered within the company and health issues. Others changed jobs in search of a better quality of life and better opportunities. Although the percentage is small, some women needed to quit their job after becoming pregnant, affecting their income and profession.

## D. Maternity Leave (RQ.2)

In spite of 93% of respondents stating that they were working when they became pregnant, only about 79% (111) of them were on maternity leave during their pregnancy. From them, 89.2% maintained the same number of employment relationships. In general, 69.4% of the mothers had one job during the maternity leave; 7.2% had two employment relationships; 2.7%, had three jobs; 5.4% more than three;

and 15.3% were unemployed. Most of them had 90 (11.7%), 120 (37.8%), or 180 (40.5%) days of license. The other 10% had more than 180 days.

Moreover, some mothers could not care for their babies full-time during maternity leave. About 24% of the respondents needed to work during maternity leave, either partially or fully. Among the reasons, were: the need for money, mentoring students, supervisor's request to return to work before the end of maternity leave, research does not take a break, they started a new job, and fear of falling behind or losing their jobs. For example, #R142 stated: *"I was afraid of losing my job and accepted my boss's blackmail to return to work after 2 months"*. However, one speech stood out: ***"My boss said I needed to prove I was capable of working after my daughter was born"***.

About 39% of respondents had no networking support during maternity leave, while 31% had partial support, for a total of 70%. We found that only 27% had some kind of support. Furthermore, 3% stated that they did not ask for any support network or that they only received help from their husband. When asked whether their partners shared childcare responsibilities with them during maternity leave, 41% said "Yes"; 32% "No", and 27% received partial help.

The support network for these mothers shortly after maternity leave was quite diverse. Figure 8 shows that the main caregivers for the children were babysitters, relatives, and nurseries, respectively. However, some mothers dedicated themselves to this task, as shown in the speech of respondent #R64: *"myself, because I always worked from home and after maternity leave came the pandemic with the beginning of the shutdown, so I didn't have much choice"*. Others had the support of their children's parents. One even reported that in Sweden parental leave is divided between both parents and added *"My husband took 6 months to leave with the child until he was 1 year old and started school"* (#R92). They also cited grandmothers, school, and mother-in-law as caregivers. Of those on maternity leave, only 13.5% indicated having support from more than one agent.

After returning to work from maternity leave, 70% of respondents work the same amount of hours per week; 17%, fewer hours per week and 13% more hours per week.

> **RQ 2 Summary**: 79% of respondents were on maternity leave during their pregnancy. Of these, 15.3% had more than one job during maternity leave and 15.3% were unemployed. Some of the women (24%) had to work during maternity leave, for a variety of reasons, among them the need for money, the fear of losing their job, or feeling less capable upon returning and being asked by their boss. About 39% of respondents had no networking support during maternity leave. The main support was provided by babysitters, family members, and nurseries.

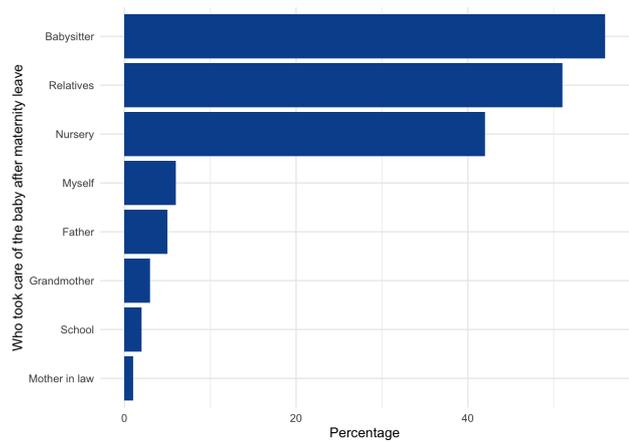

Fig. 8. Person responsible for taking care of the baby after mother's maternity leave

*E. COVID-19 Pandemic (RQ 3)*

The COVID-19 pandemic may have changed the way people relate and communicate with each other, at work and home. Several reports of women have moved from face-to-face work to remote jobs. Some companies have adopted remote work and intend to continue even after the pandemic, others are adopting a hybrid work system. It is important to highlight that this new configuration may bring benefits to the employees, however, companies should pay attention to the challenges that the configuration can bring to mothers.

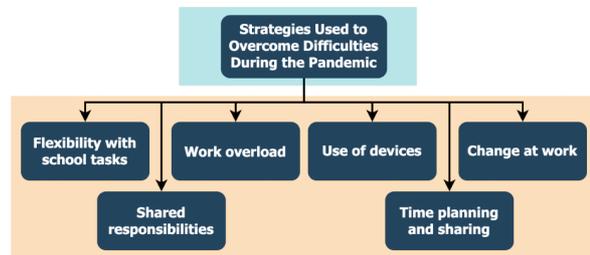

Fig. 9. Strategies used to overcome difficulties during the COVID-19 pandemic

For about 81% of respondents, the COVID-19 pandemic made it difficult to balance the time between work and motherhood-related activities. Many respondents reported strategies they employed to overcome those difficulties. We aggregated the reported strategies into six categories as shown in Figure 9. The category `Flexibility with school tasks` corresponds to strategies related to the reduction of the charge related to school activities, and the `Use of devices` represents an alternative to hold the attention of the kids while their mothers work. The other four categories contain sub-categories that are not presented in Figure 9. The `Shared responsibilities` category contains three sub-categories representing with whom they shared responsibility: some mothers were able to count on the help of *fathers or relatives*, others *hired a nanny* and others could count on *teacher assistance* to help with their children's homework.

In the `Work overload` category, a significant number of mothers reported not being able to handle their work during the day and having to work *at alternative times to the children's schedule*, such as after the child sleeps. A strategy employed by some of these mothers to overcome this difficulty is *sleep deprivation*, in other words, they slept fewer number hours per day to be able to work more.

The `Time planning and sharing` category corresponds to the creation of a plan to make it possible to work from home among the children. The respondents reported four strategies: *creating a kids schedule*, *creating a daily schedule*, *taking care of children while working*, and *taking breaks from work whenever possible*.

Finally, the last category, `Change at work` contains three strategies used by the mothers: *changing from face-to-face to home office*, *workload reduction* and *one partner quit the job* to take care of the children. From the total of responses, only one respondent did not mention any negative aspect of remote work. According to her, working remotely gave her the opportunity to spend more time with kids.

> **RQ3 Summary**: For most of the respondents (81%), the COVID-19 pandemic has made it harder to balance time between work and motherhood activities. Mothers needed to develop a set of strategies to make it possible to work during this period, such as: allowing the use of devices by the kids; workload reduction: creating a time schedule, working during alternative times to the children's schedule; and hiring a nanny or teacher assistance.

*F. Harassment (RQ4)*

We also investigated whether women had suffered any harassment because of their pregnancy. It was hoped that this type of prejudice would not occur in any proportion. However, the results pointed out that, of the 141 respondents, 59 had suffered some harassment, making up about 42%, as opposed to the other 58% (82 mothers) who had not experienced this embarrassment. Although in a smaller proportion, this percentage is quite relevant and points out that there is still an important path to be followed for actions such as these to be mitigated.

Those who suffered some harassment presented statements that reveal `gender issues` that need to be reflected upon, especially the `moral harassment` they faced. For them, the most pointed evidence were issues related to `distrust`, `mean jokes`, and related `prejudices`, as Figure 10 shows.

The `Distrust` category represents the distrust of the boss and coworkers regarding the functional capacity of the pregnant colleague or mother, as illustrated by the speeches of three of the respondents (#R143, #R130, and #R135, respectively): *"My boss said in all meetings with the team that I was not able to carry out my activities within the established deadlines due to my pregnancy"*; *"Everyone thought that my creative process would be affected by my pregnancy, that I wouldn't be creative and be able to make interesting designers"*; *"Some colleagues didn't want to be on the same team as me because they thought I wouldn't be able to work"*.

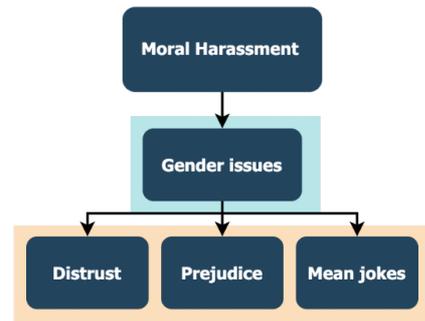

Fig. 10. Moral Harassment identified in the survey

The `Prejudice` category refers to the damages and impacts resulting from moral harassment, whether emotional, psychological, or financial. The statements brought up by the respondents shows how necessary it is to discuss gender diversity in academic and professional environments, to guarantee a more equitable, fair, and healthy space for everyone. In this category, some statements translate the cruelty of peers and, above all, the need for reflection and actions toward significant changes in organizational environments and society in general. For example, #R114 affirmed *"My colleagues always looked at me with discrimination and as if I didn't know anything about ICT"*; Also, #R131 said *"My co-workers thought that I had a disease and that I would not be able to work for the 9 months"*.

Not only colleagues, but ICT mothers also suffer from bosses' prejudice. Two statements stood out (#R115 and #R105, respectively): ***"My boss asked me to resign. He said I couldn't take on two roles, mother and practitioner"*** and ***"When I told my boss that I was pregnant, he asked the company to fire me. He said I wouldn't be able to work with kids"***.

`Mean jokes and bad jokes` were also common reports among the respondents. This category represents that these professionals, in their work environments, suffered from mean jokes throughout or after their pregnancy. For instance, #R136 affirmed *"My co-workers made a lot of jokes and started not assigning me any tasks"* and #R55 reported *"When they found out about my pregnancy they started treating me differently as if I had a very contagious disease, I always heard giggling and small talk, my paychecks were on different days than others and my boss always referred to me as 'the pregnant woman' "*.

As a consequence, due to the harassment they suffered, some women reported they began to doubt their own abilities. For instance, #R126 affirmed *"My colleagues made so many bad jokes that I began to think that I was incapable of working and being a mother"*.

> **RQ4 Summary**: 42% of the women reported that they suffered some maternity-related harassment within ICT companies/institutions. Among the most common harassment were: distrust, prejudice, and mean jokes. The women's statements illustrate the perversity of coworkers' and bosses' behaviors.

### G. Difficulties mothers face at work (RQ5)

Respondents cited many difficulties and challenges they face in their day-to-day work. For example, according to respondent #R47, it might be hard *"caring, giving attention, educating, and doing all this together with an overloaded workday"*, which can be even harder to solo mothers without a support network.

Figure 11 shows the identified categories of difficulties. Many of them are related to mothers' overload. For example, some mothers complained about the lack of time to study and `keep up to date`. The respondent #R87 said: *"One of the most difficult things is [...] not having time for training on professional skills"*. Also, there is no `time for self-care`, such as doing physical activities, mainly related to a `lack of a support network`.

Also, `balancing maternity, work, and housework` is not an easy task. Some respondents indicated that they had already missed opportunities for promotion and used to avoid complex projects to have more time for the children. They also would like to set limits on what time of the day they can have meetings, for example. A related challenge is the `lack of work flexibility` for childcare tasks.

A common difficulty among the respondents was the `maternity-related absences`. Children get sick and need to go to the doctor quite often, especially in the first years of life. The respondent #R96 said that *"[...] bosses do not understand these absences, especially those who still do not have children"*. The mothers cited five common necessary absences: medical appointments, children's illnesses, childhood vaccines, nanny absence, and school events. Another related challenge is the `lack of women` at work. The ever-reduced number of women on the team means that they have no one to talk to, which makes it difficult for colleagues and bosses to understand the difficulties. Thus, they also complained about the `lack of empathy or sorority`, sometimes even by women. The respondent #R87 affirmed: *"When I returned from maternity leave, the woman in charge gave me greater responsibilities, because when she returned from maternity leave, they did this to her"*.

`Remote work` challenges, such as children requiring attention, and the strategies to overcome them, as presented in

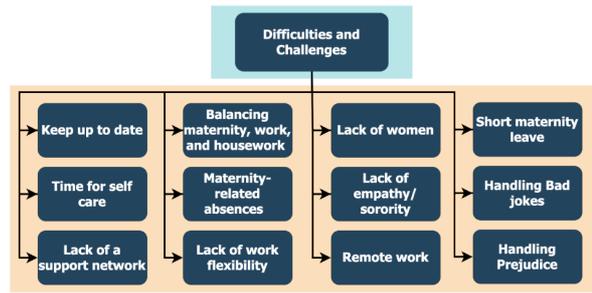

Fig. 11. Main difficulties and challenges mothers face at work

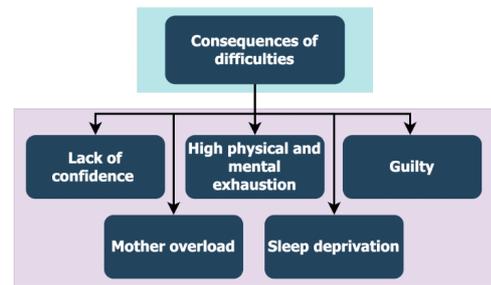

Fig. 12. Consequences of the difficulties mothers face at work

Figure 9 were also cited. Besides, the `short maternity leave` instituted in some countries, such as Brazil, can be challenging for mothers who need to return to work after 4 months, a period when the baby is still exclusively breastfed, as indicated by pediatricians.

Furthermore, mothers have to `handle bad jokes` and `prejudice` at work. For example, respondent #R136 declared "I heard many bad jokes from my male colleagues. Sometimes even women too [...]". And respondent #R107 said *"My biggest challenge is to prove that I am capable. The lack of confidence from my male colleagues makes me feel discouraged [...]"*.

The difficulties and challenges of motherhood imply a set of consequences for mothers, as Figure 12 shows. The challenges mothers face in coping with work and motherhood can have dangerous consequences. They reported being `overloaded` with `physical and mental exhaustion`, many times in `sleep deprivation`. Moreover, the `Lack of confidence` in mothers' work was a theme frequently addressed in the mothers' responses. For instance, respondent #R122 affirmed that *"Sometimes my colleagues think that I am not able to fulfill my demands at work and home"*.

They also reported having difficulty concentrating when the babies are so young and they need to go back to work because maternity leave is so short. They affirmed `feeling guilty` about: (i) being away from the children; (ii) not having time for them; and (iii) not meeting children's attention needs.

TABLE I
SUGGESTIONS TO MITIGATE THE DIFFICULTIES FACED

| ID | Suggestions | # Cited |
|---|---|---|
| S1 | Changing culture | 27 |
| S2 | Create a code of conduct for men | 21 |
| S3 | More empathy | 12 |
| S4 | Create childcare within companies | 11 |
| S5 | Create opportunities/programs for women in ICT | 11 |
| S6 | Reduce working hours | 10 |
| S7 | Hybrid work models or remote work | 7 |
| S8 | Family support | 5 |
| S9 | Network support | 5 |
| S10 | Planning, organization of maternal and work tasks | 5 |
| S11 | Avoid overtime pressure | 4 |
| S12 | Collaborative network in the company | 4 |
| S13 | Increase the sorority | 3 |
| S14 | Create gender inclusion policy/laws | 3 |
| S15 | Dialogue with children | 2 |
| S16 | Hire more women in leadership roles | 2 |
| S17 | Understanding of leaders | 2 |
| S18 | Create spaces in companies for mothers to share their difficulties and experiences (study groups) | 2 |
| S19 | Flexible hours | 2 |

> **RQ5 Summary**: Respondents pointed out many difficulties in reconciling work with motherhood. Most of them are related to mothers' overload, and the lack of time to care for themselves and keep up to date professionally. They do not find flexibility at work for the absences that motherhood requires, such as doctor appointments. Also, the bad jokes and prejudice they suffer make it harder to balance work and maternity.

*H. Suggestions to mitigate the difficulties (RQ6)*

We also asked the participants, what would be their suggestions for mitigating the difficulties (by the organization and co-workers) faced in their work environments. Table I presents all suggestions that received two or more citations.

The suggestion most mentioned by the participants was the `change of culture`, which was cited 27 times by the mothers who participated in the survey. For example, #R92 affirmed that *"The culture of the team and the company (and even the country in general) must be aligned with raising a family's difficulties. I live in Sweden and here the laws make it easier to raise children, such as parental leave of 480 days shared between the guardians of the child"*.

The second most cited suggestion was to `create a code of conduct for men`. Still, mothers suggested applying fines if men don't comply with the code, so they stop making jokes and judging the mothers' abilities. Some works in the literature proposed to develop a code of conduct as a collective policy on unacceptable behavior in interactions between members of development teams [7], [22], [24], [29], [30]. The studies report that harassment cannot be tolerated and that violations of policies should have consequences, for example by employing mechanisms to enforce the use such as appropriate penalties, if necessary.

Suggestions such as establishing limits, reducing meetings' frequency and duration, prioritizing tasks, shorter projects and less complex tasks on the return from motherhood, changing the structure of scientific events (for example, do not put submission deadlines for Sundays and holidays, have a kids' space at scientific events), making a welcome program for women, more personal days, more time off with no penalties, having professionals/colleagues who can replace mothers that need to be absent, and promote initiatives like this survey. Each of the aforementioned suggestions was cited once by the respondents.

Personal life and work should not be seen as opposing worlds, but complementary. It is essential to promote and practice conciliation strategies so that mothers do not lose their space in the labor market. Mothers need to feel welcomed in the work environment. Therefore, it is necessary to provide alternatives that promote the reception of these mothers so that motherhood is naturalized in companies. In view of what was raised, it is possible to point out some interesting initiatives capable of contributing to the transformation of organizational culture, such as:

- *Give visibility to the subject of motherhood at work*. Companies could create educational moments for leaders and other collaborators in order to raise awareness about the topic. For example, they could invite mothers to share with the entire company their challenges and how they overcame them; or solutions that would get them through.
- *Merge work groups in such a way that every team has a female member*. It is easier to deal with situations when we experience them. Including mothers in teams, in addition to increasing diversity, can make colleagues more aware of women's difficulties.
- *Provide foster care spaces for mothers*. The companies could create discussion groups for mothers, providing exchange and a safe environment. Support groups are essential for mothers to feel welcomed in the corporate environment.
- *Create spaces for children in companies and daycare centers that can receive them*. These spaces are valuable for companies, research centers, universities, and conferences.
- *Set diversity and inclusion goals*. Companies could set inclusion goals that seek to favor minority groups like working mothers. These goals need to be monitored closely. Therefore, they should be measurable and have a well-defined deadline.

**RQ 6 Summary**: The respondents feel that the work environment could be more friendly and inclusive for mothers. They suggested many actions that could be done by the organization and co-workers, such as creating a code of conduct for men and changing the culture, extending maternity leave, talking openly about maternity, and supporting parents in a way they can focus on getting their work done while they are at work.

## V. THREATS TO VALIDITY

As with any empirical study, this work also has threats to validity and limitations, which we present in this section. According to Kasunic [31], there are three important types of validity concerning to survey research: construct, internal and external validity.

The **construct validity** aims to answer if "we measuring what we think we are measuring" [31]. The questionnaire employed in this research has never been used before. In order to deal with this threat, we divide the authors into two groups: those who created the questionnaire and those who reviewed it. Furthermore, before distributing the questionnaire, we run a pilot with three people, which resulted in changes to it.

The analysis of the **external validity** of a survey aims to answer the following question "Can the results be generalized to other people, places, or times?" [31]. Despite the good number of answers (147) and the different countries that the respondents are from (17), half of the respondents of the survey are from Brazil.

Finally, about the **internal validity**, the sample characteristics might have influenced our results, for the same reason mentioned before: half of the respondents are from Brazil. Therefore, the suggestions and challenges might reflect the problems and situations related only to women that live in Brazil. Furthermore, it is important to highlight that most of the authors that conducted this research are mothers (4 out of 5) which could also have some influence on the results we got.

## VI. CONCLUSION AND FUTURE WORK

This paper presents the results of a survey about women´s perceptions of the impact of motherhood on their careers in software engineering. We received responses from 141 mothers who work with software engineering in industry and academia in 17 countries. As women and mothers, we wanted to show a more sensitive look at the difficulties of mothers working in this area and how companies can provide better support to them.

We designed a questionnaire with 55 questions to answer six research questions. The RQ 1 investigated the strategies employed by mothers to overcome difficulties during the COVID-19 pandemic. We found that most of the women (81%) reported difficulties to balance work and motherhood activities. To overcome them, they developed some strategies, such as creating a routine for the children and hiring someone to help take care of them.

The surveyed women also suggested many actions to deal with motherhood difficulties (RQ 2), such as: (i) more empathy/sorority in relation to problems related to motherhood; (ii) creating a code of conduct for men; (iii) creating childcare within companies; (iv) creating opportunities/programs for women in ICT; and (v) hybrid work models or remote work. This requires changing the company's organizational culture and placing more women in leadership positions.

Women still suffer harassment and prejudice related to motherhood (RQ 3), such as distrust and mean jokes. Some of them reported that they started to feel incapable to perform their work and being a mother, simultaneously. The RQ 4 asked how maternity leave was affected by job activities. We found out that most of the mothers (79%) could use maternity leave, however, many needed to work during it.

In relation to RQ 5 we found out that 11% of women quit after becoming mothers. Furthermore, mothers face many difficulties while working within the computing area (RQ 6), such as overload of having to comply with company work hours and overtime with household activities. They also complained that the employment makes maternity-related absences difficult, such as doctor appointments and children's illnesses.

In summary, this paper shows that mothers perceive that there is a social penalty for mothers in the corporate environment. They suffer discrimination in the area of Software Engineering, which can be very negative for the market as a whole, as it can discourage female talent and prevent their growth. Overall, everyone desires a better world to live in. A world where women and men can be treated with equality and respect, recognizing their differences. In future work, we intend to investigate how facilitating actions directed at mothers can affect factors such as team productivity and software quality. We also intend to investigate, from the perspective of men, the impact of motherhood on software development teams.